\documentclass[aps,prl,twocolumn,superscriptaddress]{revtex4}
\usepackage{amsfonts}
\usepackage{amsmath}
\usepackage{amssymb}
\usepackage{graphicx}
\usepackage{float}

\begin{document}

\title{ Folding time dependence of the motions of a molecular motor diluted inside an amorphous medium}


\author{Simona Ciobotarescu}
\affiliation{ Laboratoire de Photonique d'Angers EA 4464, Universit\' e d'Angers, Physics Department,  2 Bd Lavoisier, 49045 Angers, France}
\affiliation{ Gheorghe Asachi Technical University of Iasi, Department of Natural and Synthetic Polymers, 73, Prof. Dimitrie. Mangeron Street, 700050 Iasi, Romania}

\author{Solene Bechelli}
\affiliation{ Laboratoire de Photonique d'Angers EA 4464, Universit\' e d'Angers, Physics Department,  2 Bd Lavoisier, 49045 Angers, France}

\author{Gabriel Rajanson}
\affiliation{ Laboratoire de Photonique d'Angers EA 4464, Universit\' e d'Angers, Physics Department,  2 Bd Lavoisier, 49045 Angers, France}

\author{Samuel Migirditch}
\affiliation{ Laboratoire de Photonique d'Angers EA 4464, Universit\' e d'Angers, Physics Department,  2 Bd Lavoisier, 49045 Angers, France}
\affiliation{ Appalachian state university, Physics Department, Boone, USA}

\author{Brooke Hester}
\affiliation{ Appalachian state university, Physics Department, Boone, USA}

\author{Nicolae Hurduc}
\affiliation{ Gheorghe Asachi Technical University of Iasi, Department of Natural and Synthetic Polymers, 73, Prof. Dimitrie. Mangeron Street, 700050 Iasi, Romania}

\author{Victor Teboul}
\email{victor.teboul@univ-angers.fr}
\affiliation{ Laboratoire de Photonique d'Angers EA 4464, Universit\' e d'Angers, Physics Department,  2 Bd Lavoisier, 49045 Angers, France}

\keywords{dynamic heterogeneity,glass-transition}
\pacs{64.70.pj, 61.20.Lc, 66.30.hh}

\begin{abstract}

We investigate the dependence of the displacements of a molecular motor  embedded inside a glassy material on its folding characteristic time $\tau_{f}$.
We observe two different time regimes. For slow foldings (regime I) 
the diffusion evolves very slowly with $\tau_{f}$,  while for rapid foldings  (regime II) the diffusion increases strongly with $\tau_{f}$ (  $D\approx \tau_{f}^{-2}$) 
suggesting two different physical mechanisms. 
 We find that in regime I the motor's displacement during the folding process is counteracted by a reverse displacement during the unfolding, while in regime II this counteraction is much weaker. We notice that regime I behavior is reminiscent of the scallop theorem that holds for larger motors in a continuous medium.
We find that the difference in the efficiency of the motor's motion explains most of the observed difference between the two regimes.
For fast foldings the motor trajectories differ significantly from the opposite trajectories induced by the following unfolding process, resulting in a more efficient global motion than for slow foldings.
This result agrees with the fluctuation theorems expectation for time reversal mechanisms. 
In agreement with the fluctuation theorems we find that the motors are unexpectedly more efficient when they are generating more entropy, a result that can be used to increase dramatically the motor's motion.

\end{abstract}

\maketitle
\section{ Introduction}
The development of molecular motors have received a large attention\cite{motor1,motor2,motor3,motor4,motor5,motor6,motor7,motor8,motor10,motor11,motor12,motor13,motor14,motor15,motor16,motor17,pccp}
 since the beginning of nanotechnology. 
 Molecular motors can be designed to move inside liquids, viscous media or soft matter. 
 
Inside supercooled liquids and soft matter their motion is even more interesting as a source of information on the still unsolved physics of  
the glass-transition\cite{anderson,gt1,gt4,gt2,adam,chandler,frust,frust2,RFOT,RFOT2,MCT,defect,facilitation,dh1,dh2,dh3,dh4,dh5,fragile1,fragile2}.
Indeed, characteristics of the glass-transition physics like cooperative motions called dynamic heterogeneities were reported to be generated during the motion of various motors\cite{prl,us2,c1,c2,c3}.
 The photo-fluidization and softening of the host material, or transient liquid-like behaviors,  were also reported by different groups experimentally and by simulations\cite{dif1,dif2,dif3,dif4,prl} with particular motors.
These behaviors and possible cage-breaking processes induced by the motor\cite{cage,c5} suggest that molecular motors small stimuli can be used to probe the physics of the glass-transition.

The problem of the motion of molecular motors is complicated by the existence of Brownian motion that washes out any attempt to move constructively at the nanoscale.
Molecules like stilbene, azobenzene and their derivatives, do  have the property to fold when illuminated due to a photo-isomerization process, and are of particular interest as motors because they do not consume or produce any waste inside the host medium.
When illuminated, azobenzene doped materials are subject to intriguing macroscopic transport that in some conditions lead to the formation of surface relief gratings (SRG)\cite{review}.
While the exact physical mechanisms leading to that macroscopic transport is still a matter of debate\cite{review,review2,review4,a17,a18,a21,a22,a23,a24}, there is no doubt that it originates from the repeated foldings of the azobenzene molecule inside the material.

The characteristic times involved in the molecular motor's folding are of importance as we may expect them to control the physical mechanisms behind the medium and motor's motions.
In a previous paper\cite{rate} we have studied the effect of the folding rate $1/T_{f}$ on the molecular mobilities, and found that for small rates the mobilities followed the linear response theory, then saturate at larger rates. 
In this work we study the effect of the second characteristic time, the motor molecule folding time $\tau_{f}$, on the motor's motion.
If the folding rate $1/T_{f}$ controls the amount of energy that is released inside the medium per unit of time, the folding time $\tau_{f}$ controls the forces created during the folding process.
We expect the molecular motor motions to increase with the folding forces thus to increase when $\tau_{f}$ decreases.
Results show the presence of two different dynamical regimes suggesting different physical mechanisms. For rapid foldings the motor and medium motions are proportional to the force created by the motor's folding. In that dynamical regime the motor's motion increase with the force as $\tau_{f}^{-2}$.  For applicative purpose, a variation of the folding time in that regime could increase importantly the motor and host motions. We note that in this regime, our results agree remarkably well with the gradient pressure theory\cite{a22,a23} for the formation of SRG. 
Then the system reaches a different dynamical regime when slow foldings are used. In that regime the motor and medium motions only slightly depend on the folding time $\tau_{f}$ and thus do not depend significantly on the force.

Our simulations show in this paper that for most foldings in regime I the motor's displacement during the folding process is hindered by a reverse displacement during the unfolding, but not in regime II. 
To quantify this effect that is reminiscent of Purcell's scallop theorem\cite{scallop1,scallop2,scallop3}, we define the efficiency of mobility $\epsilon$ as the average motion of the motor during $p$ foldings and $p$ unfoldings (i.e. $p$ periods), divided by $2p$ times  its average motion for one folding only:

\begin{equation}
\epsilon=<r^{2}(p T_{f})>/(2p<r^{2}(T_{f}/2)>)  \label{e1}
\end{equation}

With this definition $\epsilon=0$ for a totally inefficient process where the unfolding motion of the motor is the reverse of its folding motion, while $\epsilon=1$ for an efficient process for which the unfolding motion is not correlated with the folding motion. Note that values of $\epsilon$ larger than $1$ while unlikely, are possible with this definition.

We find that the difference in the efficiency of mobility explains most of the observed difference between the motions in the two regimes.
In other words, for fast foldings the motor trajectories differ significantly from the opposite trajectories induced by the following unfolding process, resulting in a more efficient global motion than for slow foldings.
This result agrees well with the fluctuation theorems\cite{Crooks,Crooks1,Crooks2,Crooks3} expectation for time reversal mechanisms.

The scallop theorem\cite{scallop1} stands that due to the very small Reynolds numbers associated with small systems, the hydrodynamics equations are approximately symmetric in time and as a result swimming using symmetric reverse motions is not possible.
While we are at the nanoscale, and as a result not in a continuous medium governed by the hydrodynamics equations, that theorem shows anyway a tendency that will become more and more correct as the motor is made larger.
Interestingly, even for larger scales where the theorem is applicable, various cases violating the theorem have been reported\cite{scallop2,scallop3}.
Two important cases violating the theorem are non-Newtonian fluids\cite{scallop2} and systems undergoing fluctuations\cite{scallop3}.

Similarly, at the nanoscale, the fluctuation theorems\cite{Crooks,Crooks1,Crooks2,Crooks3} quantify the probability that a reverse trajectory takes place in relation with the entropy generated. Crook's fluctuation theorem\cite{Crooks} stands that the probability of a reverse trajectory is proportional to $exp(S/k_{B})=exp(W_{d})/k_{B}T$ where $W_{d}$ is the amount of work dissipated in the forward trajectory $W_{d}=(W(\tau_{f})-W(\tau_{f}=\infty))$ assuming that  $W(\tau_{f}=\infty)=\Delta F$ is the reversible part of the work i.e. the difference between the free energies after and before the motor's motion.  
The fluctuation theorems thus show that, when out of equilibrium, reverse trajectories do not have the same probability than the direct trajectories, showing that the motor's motion is possible at the nanoscale. The reverse trajectories hindering the motions in the scallop theorem, are at the nanoscale less and less probable when the entropy generated is made larger.
That result is in qualitative agreement with our findings. 
The motion of the motor at the nanoscale requires thus to be efficient that some entropy is generated.
Assuming that the dissipated work is proportional to the force induced by the motor, the motions of pushed surrounding molecules being limited by the size of the motor's arms, the Crook's theorem leads in the limit of small entropies to a $\tau_{f}^{-2}$ evolution of the efficiency of mobility in agreement with regime II.
Following this viewpoint, a tentative explanation of regime I is that it arises due to the minimum work that has to be dissipated to break the cage of the neighbors in our supercooled medium, for the folding and unfolding to take place.

\section{Model}

Molecular dynamics and Monte Carlo simulations methods are important tools to unravel the physics of materials at the atomic scale\cite{mdcm-1,mdcm-2,md7,mdcm-3,mdcm-4,mdcm-5}.
We simulate the folding of a molecular motor molecule inside a medium composed of 500 methylmethacrylate ($C_{5}H_{8}O_{2}$) molecules (note that the molecule is not polymerized).
We use that molecule as a model system in this work, chosen because of its well established characteristics from a number of previous works.
A detailed description of the simulation procedure can be found in previous works\cite{prl,rate,cage,pccp,angle}.  We model the medium molecule with the  $4$ centers of force coarse-grain potential function described in ref.\cite{md4}. The mass of the medium molecule is $m=100g/mole$. 
The  molecular motor and the associated potential functions are described in ref.\cite{pccp} and shown in Figure 1.
It is constituted of two parallel ranks of 7 atoms distant of $2$\AA\ in the rank and of $1$\AA\ between two ranks.
Each of the atoms has a mass of $40g/mole$ and is described by a Lennard-Jones 6-12 potential with $\epsilon=1 KJ/mole$ and $\sigma=3.4$\AA. 
The motor's mass is  $M=560 g/mole$.
The density is set constant at  $\rho=1.65 g/cm^{3}$. This relatively large density was chosen to increase the viscosity of the material. We evaluate the glass transition temperature to be $T_{g}=550 K$ in our system. Consequently at the temperature of the simulations $T=140$ K, there is no visible thermal diffusion ($D_{th}\approx 0$) in our simulations when the motor is inactive. 
Our cubic simulation box contains $N=2014$ centers of force (describing $7514$ atoms) and is $37$ \AA\ wide. 
A few simulations with a larger box at the same density, containing $2000$  medium molecules that is $N=8014$  centers of forces,  within a box  $58.7$  \AA\ wide, insured us that no size effects nor indirect interactions with replica of the motor are visible in our data. To highlight that point we show in Figure 2A  motor's mobilities obtained with $N=8014$ centers of forces together with results using $N=2014$ for comparison.
We use the Gear algorithm with the quaternion method \cite{allen,md2,md2b} to solve the equations of motions with a $\delta t=10^{-15} s$ time step. The temperature is controlled using a Berendsen thermostat \cite{berendsen}.

\begin{figure}[H]
\centering
\includegraphics[height=5.5cm]{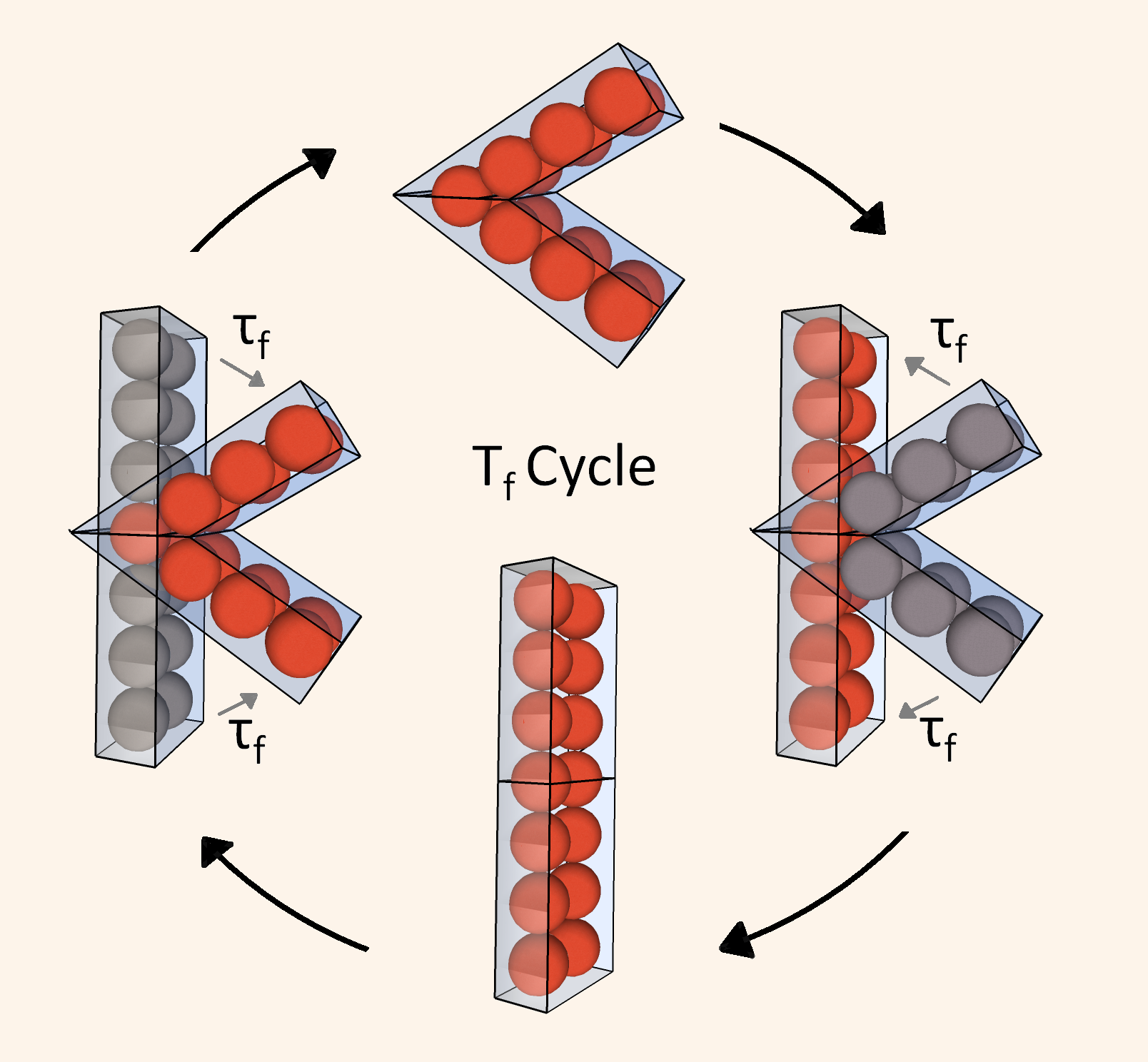}

\caption{(color online) Picture of the cycle showing the folding and unfolding of the motor's molecule. The total period of one cycle is $T_{f}$ while $\tau_{f}$ is the folding time. The motor is described with 14 centers of forces shown as atoms on the picture. The  parallelograms are guides to the eye.
} 
\label{f0}
\end{figure}

We model the folding process as a uniform closing and opening of the probe molecule shape during a characteristic time $\tau_{f}$ with a period $T_{f}$.
The motor folds in a time $\tau_{f}$ then stays folded during a time $(T_{f}/2-\tau_{f})$ then unfolds in a time $\tau_{f}$, stays unfolded during a time $(T_{f}/2-\tau_{f})$ and the cycle continues.
When unfolded the molecule is a planar rectangle of length $L=15.4$ \AA\ and $l=4.4$ \AA\ wide (see Figure 1).
After the folding the molecule is folded on an axis passing through its mid length, with an angle $\alpha=60$ degrees. 
$T_{f}/2$ appears as the time lapse between two energetic impulses inside the material and will be an important characteristic time in our study.
However the effect of a variation of $T_{f}$ has been previously studied\cite{rate} and we will focus in this work on the other important characteristic time $\tau_{f}$.

There is only one motor inside the simulation box.  
We use this small motor's concentration to have, in our simulations, the smallest possible perturbations of the host material for a given folding period $T_{f}$. We compare the results with two different folding periods $T_{f}=400 ps$ and  $T_{f}=600 ps$ to verify that we are in the linear response regime\cite{David} i.e. that the response is proportional to the perturbation ($D =\alpha/ T_{f}+ D_{th}=\alpha/ T_{f}$).
This regime insures us that the foldings do not interfere with the following ones, i.e. each folding acts on an unperturbed medium and thus can be seen as unique.
We display the diffusive behavior of the motor and host's motions in the appendix.

\section{Results and discussion}

\subsection{Diffusion of the motor and host versus folding time: Two time regimes}

\begin{figure}[H]
\centering
\includegraphics[height=7cm]{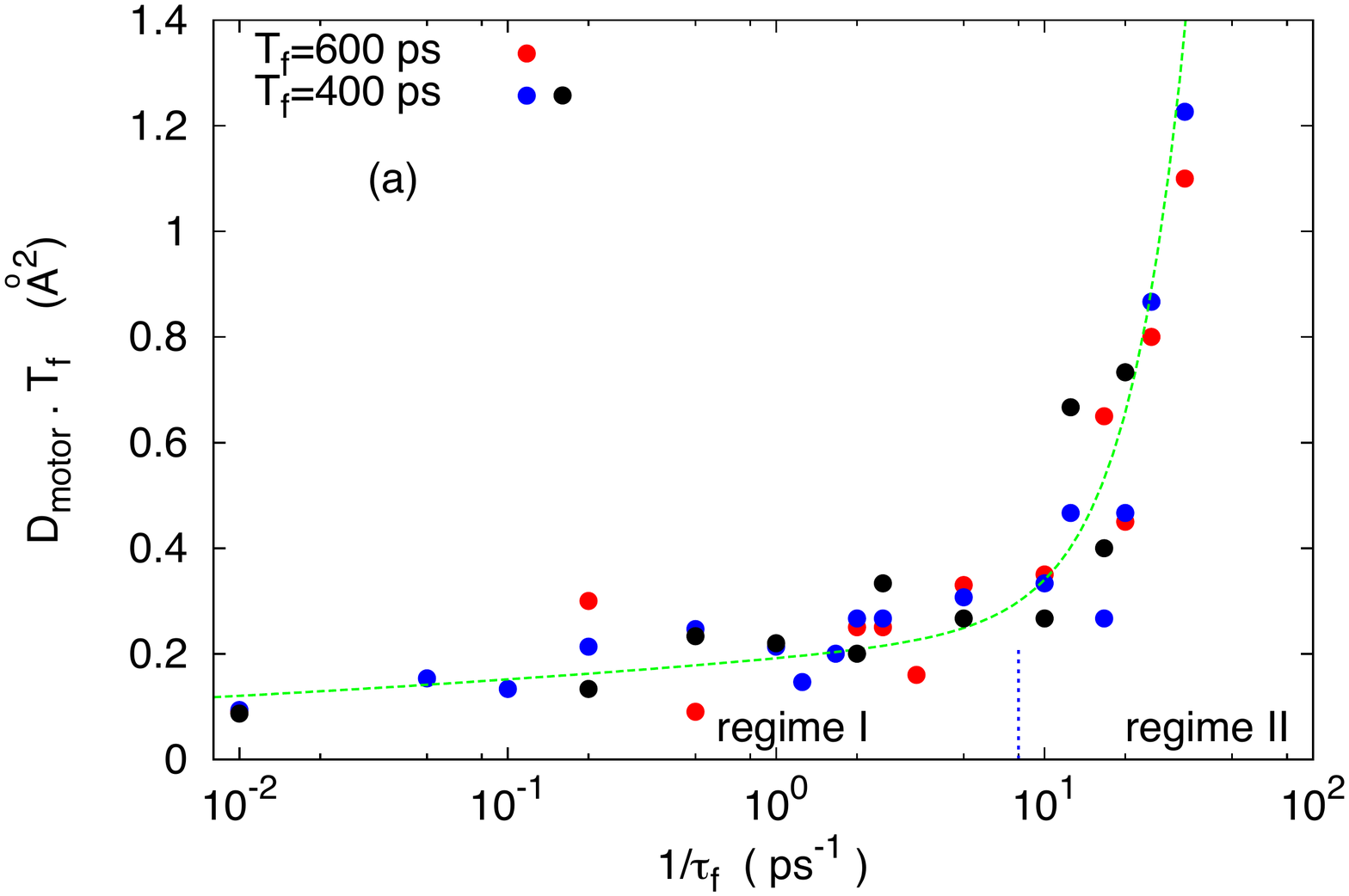}

\includegraphics[height=7cm]{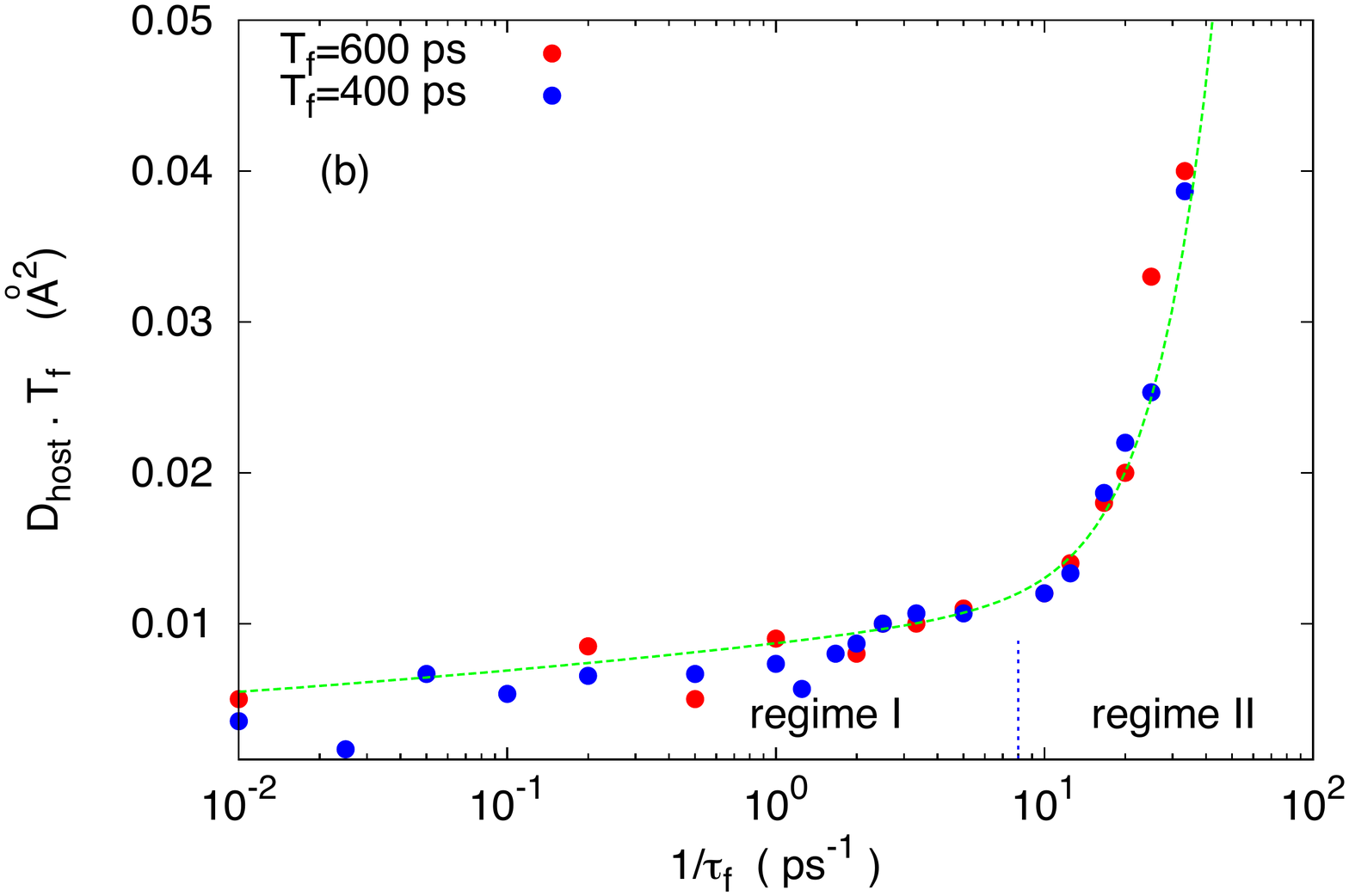}

\caption{(color online) (a) Motor's diffusion coefficient $D_{motor}$ multiplied by the folding period $T_{f}$ versus the inverse of the folding time $\tau_{f}$. Blue (light), red (gray) and black (dark) circles correspond to two different folding periods.
Black circles correspond to large simulation boxes. The small blue dotted line corresponds to $\tau_{f}=1.25$ $10^{-1} ps$ and separates the two regimes. The temperature is $T=140 K$. The green dashed line is a fit corresponding to $D_{motor}=a \tau_{f}^{-0.1}+b \tau_{f}^{-2}$ (as $T_{f}$ is constant). Note that the data are also compatible with a fit corresponding to $D_{motor}=c + d \tau_{f}^{-2}$; $a, b, c, d$ are  constants that depend on the motor. 
(b) Same Figure but for the  host averaged diffusion $D_{host}$.
The green line fit corresponds to $D_{host}=e \tau_{f}^{-0.1}+ f \tau_{f}^{-2}$ where $e$ and $f$ are constants.
} 
\label{f1}
\end{figure}

 Figure \ref{f1} shows the evolution of the diffusion of the motor and host with the folding time $\tau_{f}$ at low temperature.
The red and blue circles represent two different folding periods ($T_{f}=600ps$ and $T_{f}=400ps$) while the black circles correspond to large simulation boxes with $T_{f}=400ps$. The different set of points merge, showing that we are in the linear response regime,  as the response is here proportional to the number of foldings per second, i.e. to the stimulus. 
This behavior suggests that cumulative effects  on the medium due to the periodic foldings do not affect significantly our results.
We observe on both figures two different folding time regimes. For slow folding times (regime I) the diffusion coefficient evolves slowly with $\tau_{f}$ ($D_{motor}\approx \tau_{f}^{-0.1}$), while for fast folding times ($\tau_{f}<1.2$ $10^{-1} ps$, regime II) the diffusion increases strongly with $1/\tau_{f}$ ($D_{motor}\approx \tau_{f}^{-2}$).
These differences suggest two different physical mechanisms.
We estimate the variation of the momentum during the folding process $\delta p \approx n m \delta v$, with $\delta v \approx L/(2\tau_{f})$, where $n$ is the number of host molecules pushed by the motor during its folding and $m$ is the mass of a host molecule.
Consequently the instantaneous force generated by the medium on the motor is approximately: $F \approx \delta p/\tau_{f} \approx n m L^{}/(2. \tau_{f}^{2})$.
This force evolves as $\tau_{f}^{-2}$ like the motor's motion in regime II.
This result suggests that regime II is dominated by the forces generated by and on the motor.
Regime I slow evolution is more surprising. We see on the figures that the motions still take place in that regime for very long folding times. 
Notice however that we are at low temperature and  thermal diffusive motions are small.
The figure also  shows that the host follows the same trend than the motor, the motor being faster by a rough factor $4$.
That result suggests that in our simulations the host motions are driven by the motor's motions.

\subsection{Forces on the motor}

\begin{figure}
\centering
\includegraphics[height=7cm]{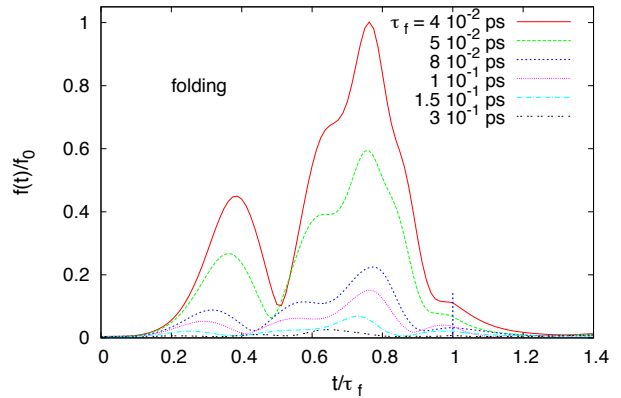}
\caption{(color online) Time evolution of the force acting on the motor during the folding process for various folding times $\tau_{f}$.
The unfolding time evolution  is quite similar.
The force persists during a short time after the end of the folding (small blue dashed line), due to the perturbation of the host medium. 
} 
\label{force1}
\end{figure}

Figure \ref{force1} shows the force evolution on the motor during the folding process.
The force evolution is not monotonous but there are two peaks with a minimum around the half folding time.
As the folding molecule probes the surrounding structure of the medium it finds larger resistances  when it encounters the surrounding molecules shell.
As a result that force evolution is characteristic of the structure of the medium surrounding the probe. 
Figure  \ref{force1}  shows that the maximum of the force arises during the last part of the folding.

\begin{figure}
\centering
\includegraphics[height=7cm]{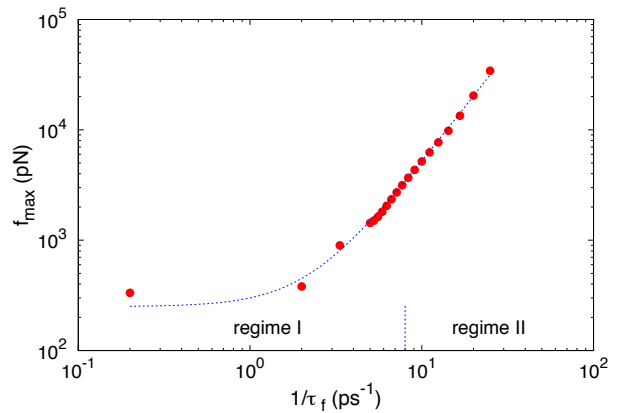}
\caption{(color online) Maximum value of the force acting on the motor during the folding process versus folding characteristic time $\tau_{f}$.
The maximum value of the force evolves as $\tau_{f}^{-2}$  but a departure from this law arise for slow foldings (first points on the curve: $\tau_{f}>200 ps$ i.e. regime I) as there is a minimum force requested to push away the surrounding hosts to permit the motor's folding. The dashed blue line is a fit of the form $f=a+b \tau_{f}^{-2}$.
} 
\label{force2}
\end{figure}

We find in Figure \ref{force2} that the maximum of the force evolves as $ \tau_{f}^{-2}$, a result that confirms the simple evaluation of the force discussed in the previous section.
This result implies that in regime II the diffusion is proportional to the maximum of the force.
It suggests that in regime II, the maximum of the force induced by the folding of the motor is at the origin of the motor and host motions. 
Regime II results thus agree with the gradient pressure model\cite{a22,a23} proposed to explain the azobenzene isomerization-induced massive mass transport. 
In regime I, the motor and hosts motions depend only slightly on the force created on the environment during the folding.
That result agrees with models like the caterpillar motions model\cite{a24} and the cage breaking model\cite{cage} as both models depend only slightly on the force generated by the motor. In the caterpillar model the motor's motion is generated by the slithering of the motor inside the medium and thus needs only small forces. Similarly in the cage breaking model the motion is generated by the breaking of the cage and the motor thus only needs the force to break one cage at each folding.

\vskip 1cm

\subsection{Efficiency of mobility in regimes I and II}

\begin{figure}[H]
\centering
\includegraphics[height=7cm]{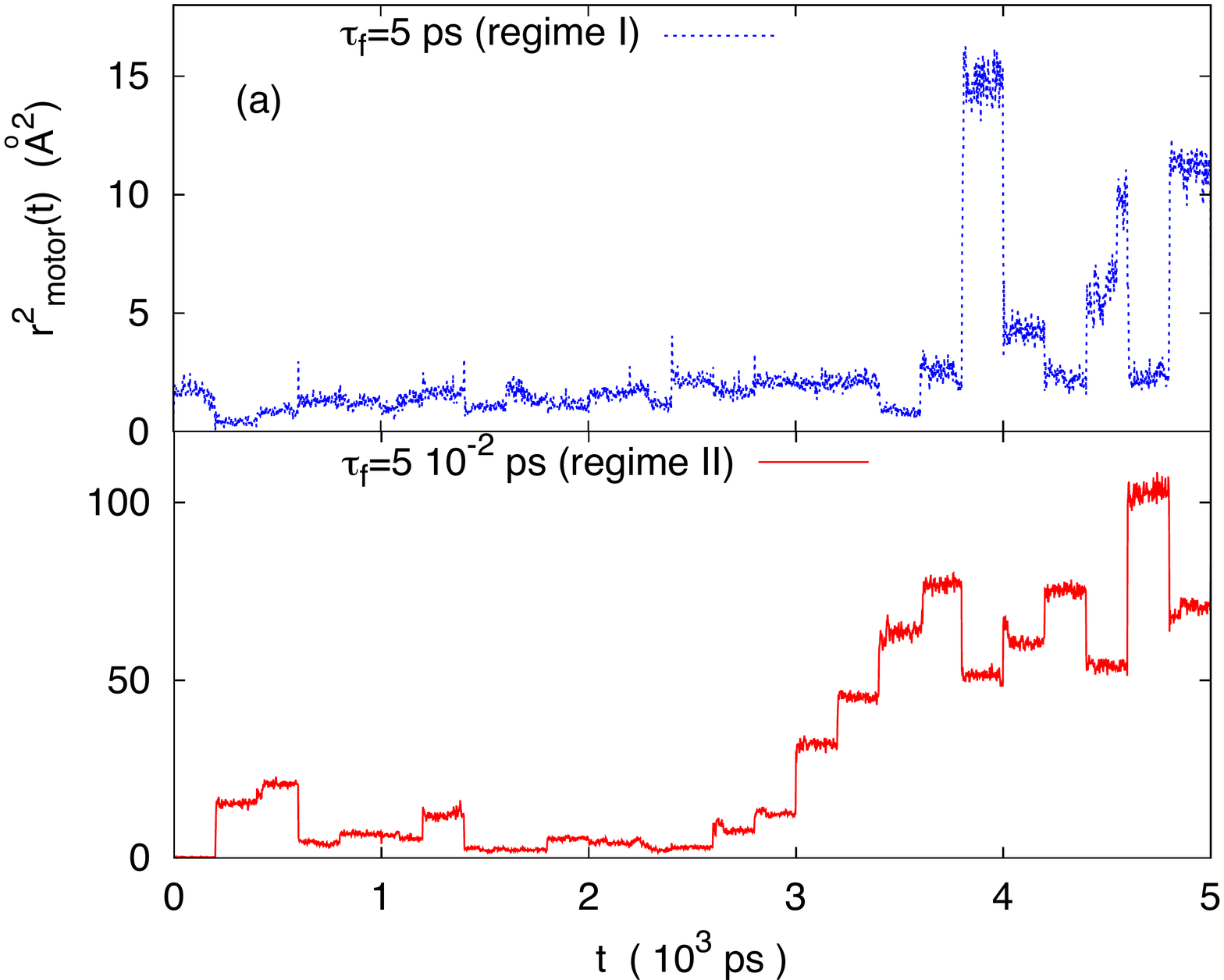}

\includegraphics[height=7cm]{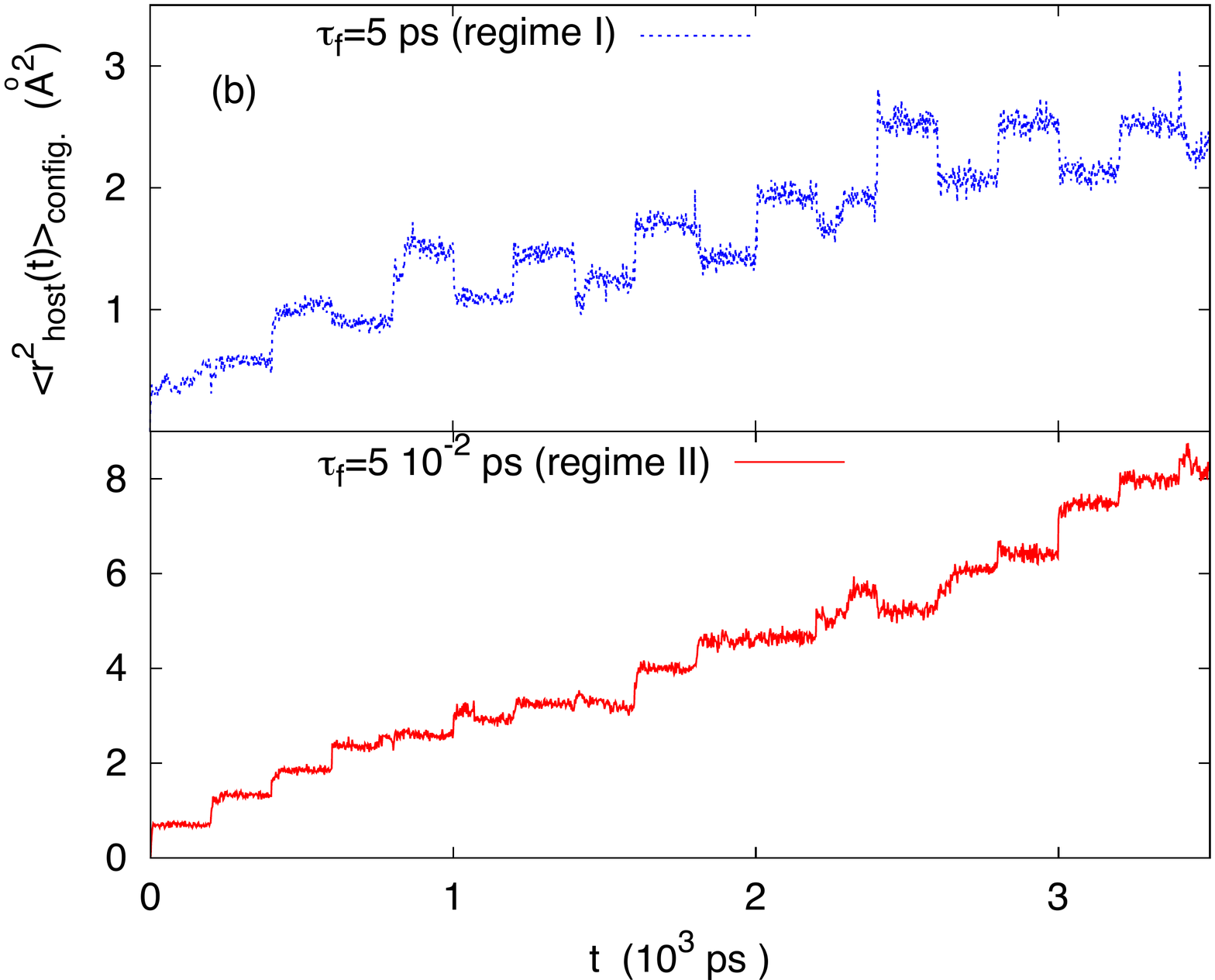}

\caption{(color online) (a) Square displacement of the motor ${r^{2}(t)}$, versus time $t$. Red dark line:  regime II ($\tau_{f}=5$ $10^{-2} ps$), the square displacement is here rescaled by a factor $1/5$ for better clarity of the Figure; Blue light line: regime I ($\tau_{f}=5 ps$). These data are not averaged on time origins. The period is $T_{f}=400 ps$.
(b) Configurational average of the square displacement of the host molecules $<r^{2}(t)>_{config.}$ located around the motor (at a distance $R<10$\AA), versus time $t$. Red dark line:  regime II ($\tau_{f}=5$ $10^{-2} ps$) rescaled by a factor $0.5$; Blue light line: regime I ($\tau_{f}=5 ps$). These data are not averaged on time origins, but are averaged on  medium molecules. $T_{f}=400 ps$.} 
 \label{msd1}
\end{figure}

\begin{figure}[H]
\centering
\includegraphics[height=7cm]{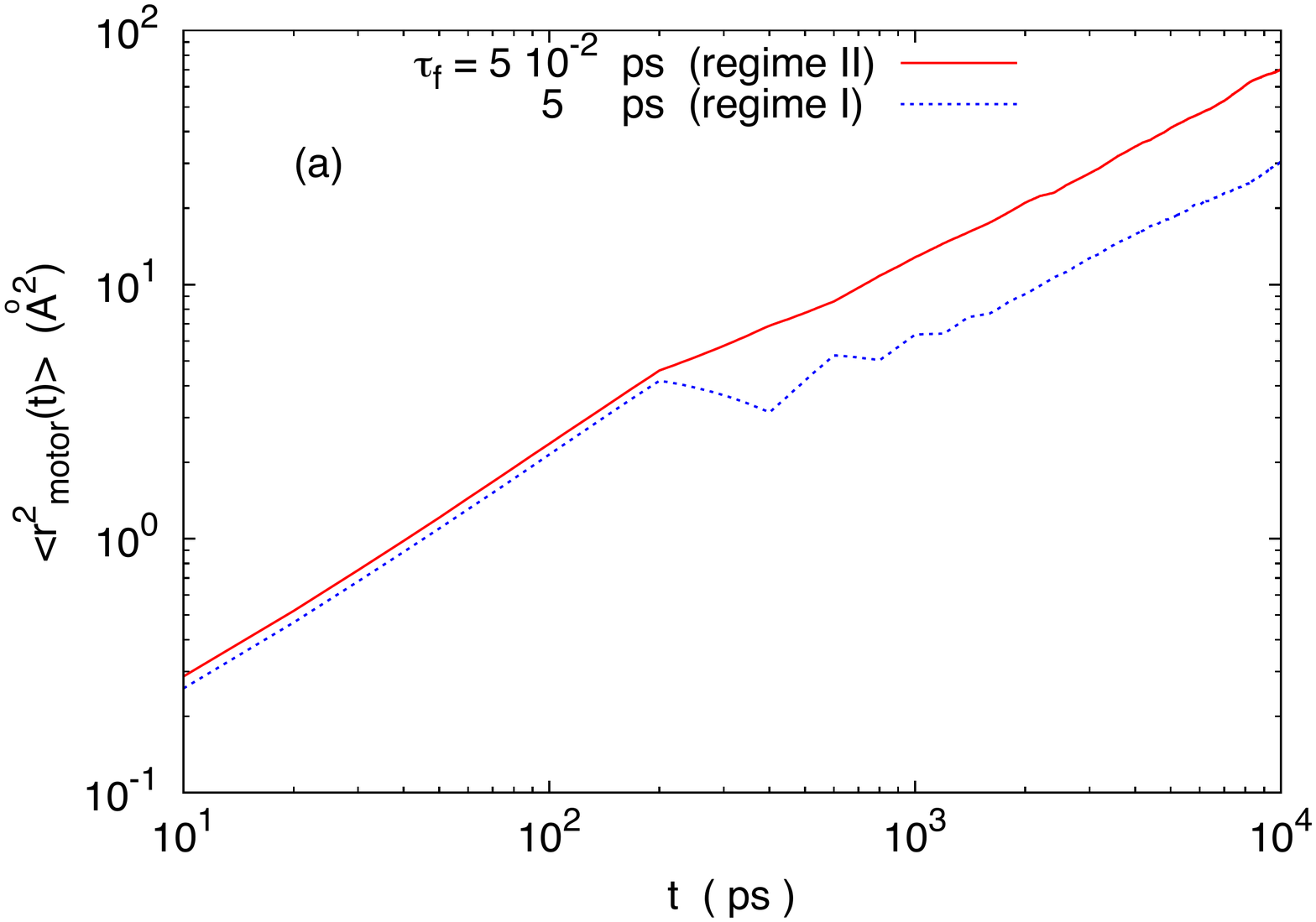}

\includegraphics[height=7cm]{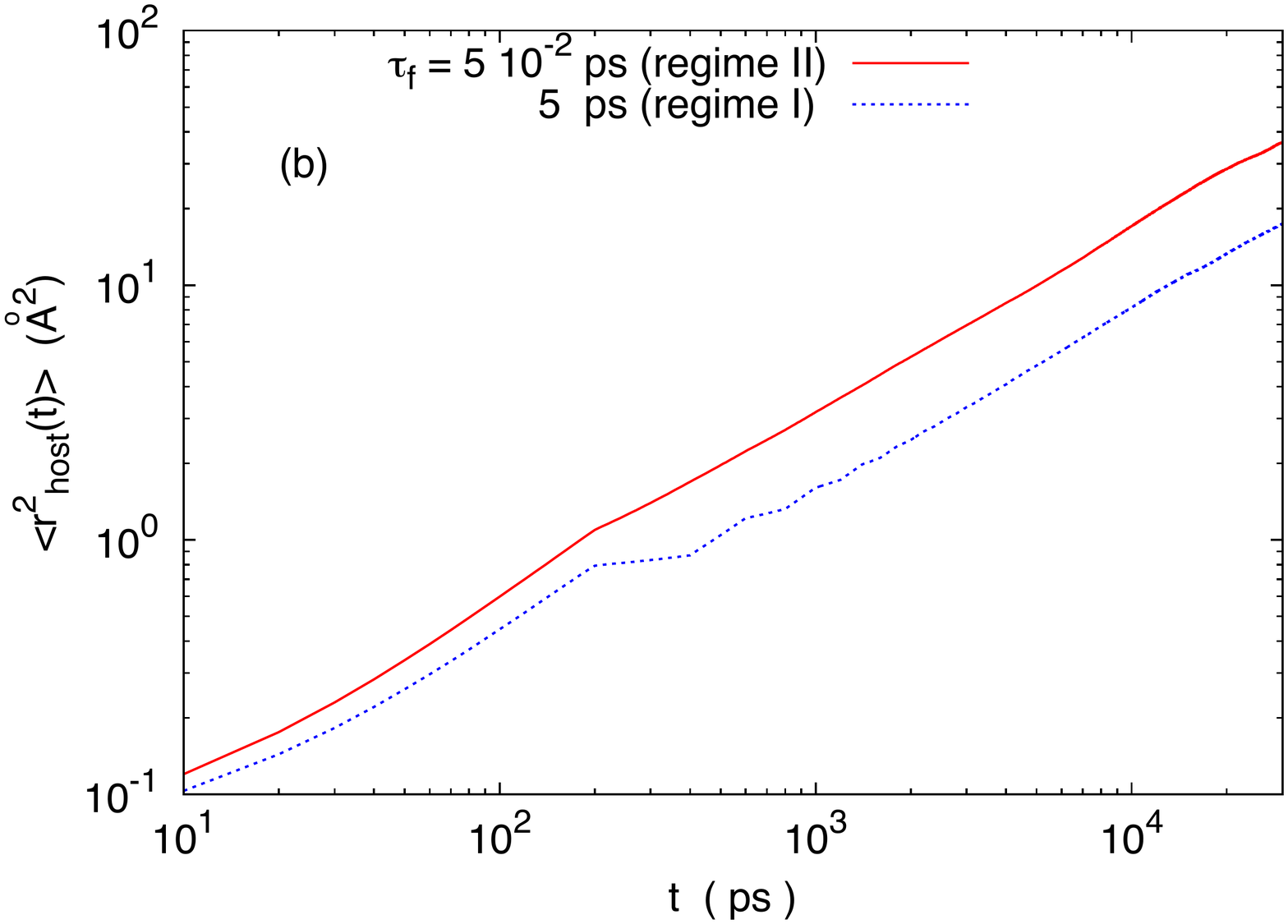}

\caption{(color online) (a) Mean square displacement of the motor $<r^{2}_{motor}(t)>$, versus time $t$. Continuous red dark line:  regime II ($\tau_{f}=5$ $10^{-2} ps$); Dashed blue light line: regime I ($\tau_{f}=5 ps$). These data are averaged on time origins.
The two curves are almost identical up to $t=T_{f}/2=200 ps$ (i.e. the first peak of the dashed blue curve, that is the time of the 'first folding') then the two curves separate due to a decrease of the light blue curve (regime I) during the following 'unfolding' while the  red continuous curve (regime II) continues to increase.
This result shows that the unfolding motor's motion opposes the folding motion for regime I and not for regime II.
Note that because $<r^{2}_{motor}(t)>$ is averaged on the time origins, the same conclusion will arise if we replace 'folding' by 'unfolding'.
(b) Mean square displacement of the host molecules $<r^{2}_{host}(t)>$ located around the motor (at a distance $R<10$\AA), versus time $t$. Red line:  regime II ($\tau_{f}=5$ $10^{-2} ps$); Blue line: regime I ($\tau_{f}=5 ps$). These data are averaged on time origins. We observe the same trend than for the motor in (a).
} 

 \label{msdt1}
\end{figure}

To better understand the origins of the two regimes, we plot in Figure  \ref{msd1}  the square displacement of the motor and mean square displacement of the hosts versus time.
The Figures show the periodic displacements induced by the folding and then unfolding of the motor.
The motor's and host displacements within time regime I (here $\tau_{f}=5 ps$, blue curves) are mostly inefficients as unfolding motions oppose the folding motions, leading to the rectangular displacements steps observed in Figure  \ref{msd1}. The long time displacement are here due to a few steps that do not come back to their exact positions during the unfolding process. 
On the contrary in regime II (red curve, $\tau_{f}=5$ $10^{-2} ps$),  most steps are efficient leading to fast diffusion of the motor and host.
We find that some motor's steps are quite high in regime II while the host steps are roughly constants.
These results confirm that the motor's motion drives the host dynamics, and show that the slow dynamics of regime I is partly due to the inability of the motor to reach steps large enough to modify significantly the return direction during the unfolding process.
The mean square displacements in Figure  \ref{msdt1}  that are averaged on the whole set of time origins, confirm that interpretation. 
The Figures show that in regimes I and II, the mean displacements are quite similar for short time scales. 
Up to $200$ ps (i.e. $T_{f}/2$) the blue curves and the red curves follow the same increase in Figure \ref{msdt1}.
Then the red curve (regime II) continues to increase while the blue curve (regime I) oscillations corresponding to the unfolding and folding times result in a smaller increase with time. In other words the elementary displacement (i.e. corresponding to one folding only) is the same in regimes I and II, while the long time displacements are not. 
This result suggests that the difference between regime I and II arises mainly from a difference in the efficiency of mobility of the motor (i.e. the probability that the unfolding process doesn't reverse the motion induced by the folding) and only more slightly on the forces pushing the motor forward. 

\begin{figure}[H]
\centering
\includegraphics[height=7cm]{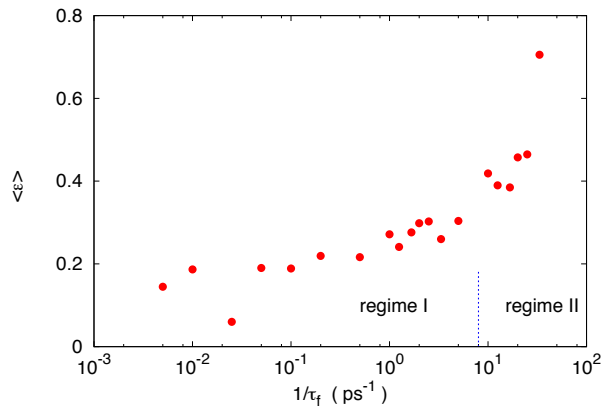}

\caption{(color online) efficiency of mobility of the motor defined as $\epsilon=<r^{2}(p T_{f})>/(2p<r^{2}(T_{f}/2)>)$ versus the folding time $\tau_{f}$. For the values plotted $p=5$ and $T_{f}=400 ps$. 
An efficiency of mobility of $1$ means that the motor has the same probability to return to its initial position than in a Brownian random motion, while an efficiency of mobility of $0$ means that the motor always return to its initial position (i.e. that the unfolding process destroys the motion induced by the folding). 
} 

 \label{eff}
\end{figure}

We will now use the efficiency of mobility $\epsilon$ defined in the introduction (equation 1) to quantify the hindering of the motor's motion due to the lack of efficiency of mobility discussed previously and compare its evolution with the motion induced by the folding.
The evolution of $\epsilon$ with the folding time $\tau_{f}$ is displayed in Figure \ref{eff}.
The Figure shows that the efficiency of mobility $\epsilon$ increases with $1/\tau_{f}$ leading to a much larger efficiency of mobility in regime II than in regime I.
The increase in the efficiency of mobility thus can explain the difference in the motor's motion between the two regimes.
We will now show that the change in the efficiency of mobility is indeed the main contribution to that difference of motions between the two regimes.

\begin{figure}[H]
\centering
\includegraphics[height=7cm]{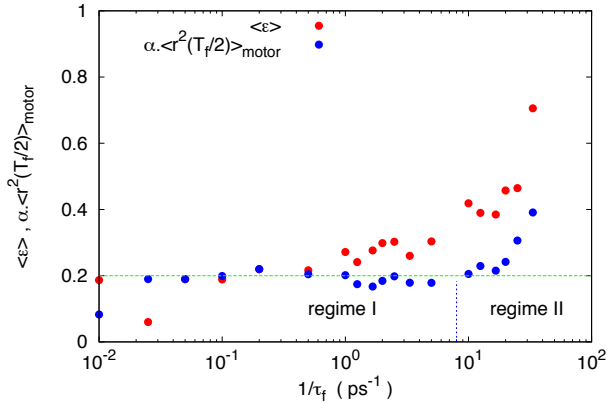}

\caption{(color online) Comparison between the efficiency of mobility $\epsilon$ evolution versus the folding time $\tau_{f}$ and the motor's motion after one folding only $<r^{2}(T_{f}/2)>$.
The mean square displacement $<r^{2}(T_{f}/2)>$ is multiplied by a factor $\alpha=5.3$  $10^{-2} $ \AA$^{-2}$ to help compare the two evolutions. 
The green dashed line is a guide to the eye showing that the mean square displacement of the motor for one folding is approximately constant up to $1/\tau_{f}=20 ps^{-1}$ that is for $\tau_{f} \geqslant 5$ $10^{-2} ps$. } 

 \label{effn}
\end{figure}

We show in Figure \ref{effn} the evolution of the average motor's motion after the folding process and we compare its evolution with the evolution of the efficiency of mobility in order to understand which one of these two contributions is the main cause of the motion's evolution with $1/\tau_{f}$.
Figure \ref{effn} shows that the average motor's motion after one folding also increases with $1/\tau_{f}$ although relatively slower than the efficiency of mobility.
The motor's motion during one folding process follows a similar trend than the efficiency of mobility and global motion and contribute to the increase in the motor's motion with $1/\tau_{f}$.
The comparison between these two contributions in Figure \ref{effn} shows that the efficiency of mobility has a larger contribution to the global motion evolution.

 Figures   \ref{eff}  and \ref{effn} show that the difference in the efficiency of mobility explains most of the observed difference between the motions in the two regimes.
For fast foldings the motor trajectories differ significantly from the opposite trajectories induced by the following unfolding process, resulting in a more efficient global motion than for slow foldings.
As discussed before, this result agrees with the fluctuation theorems\cite{Crooks,Crooks1,Crooks2,Crooks3} expectation for time reversal mechanisms.

\section{Conclusion}

In this work we have studied the effect of its characteristic folding time  $\tau_{f}$ on a molecular motor's displacements inside a soft material well below its glass transition temperature $T_{g}$. 
We found two different dynamical regimes.
For slow foldings (regime I) the motor's motions are only weakly dependent on the folding time ($D_{motor} \approx \tau_{f}^{-0.1}$).
For rapid foldings (regime II) in contrast, the motor's motions strongly depend on the characteristic folding time ( $D_{motor}\approx \tau_{f}^{-2}$) as the maximum value of the force induced by the folding on the motor.
We found that the difference between the two regimes mainly arise due to a different efficiency of mobility of the motor.
For slow foldings the unfolding process destroys most of the motion induced by the folding, while this is not happening for rapid foldings.
We interpret that difference as arising from the increase of irreversibility for rapid foldings.
When the folding is fast enough, it changes significantly the motor's environment leading to an irreversible and more efficient process.
Experimentally the characteristic folding times can be modified by acting on the motor's molecule electronic structure and to a lesser extent on the viscosity of the environment. 
Our results show that above a threshold value, rapid characteristic folding times can dramatically increase the motions of a molecular motor in soft matter.

\section{Appendix: Diffusive motions}

\begin{figure}[H]
\centering
\includegraphics[height=7cm]{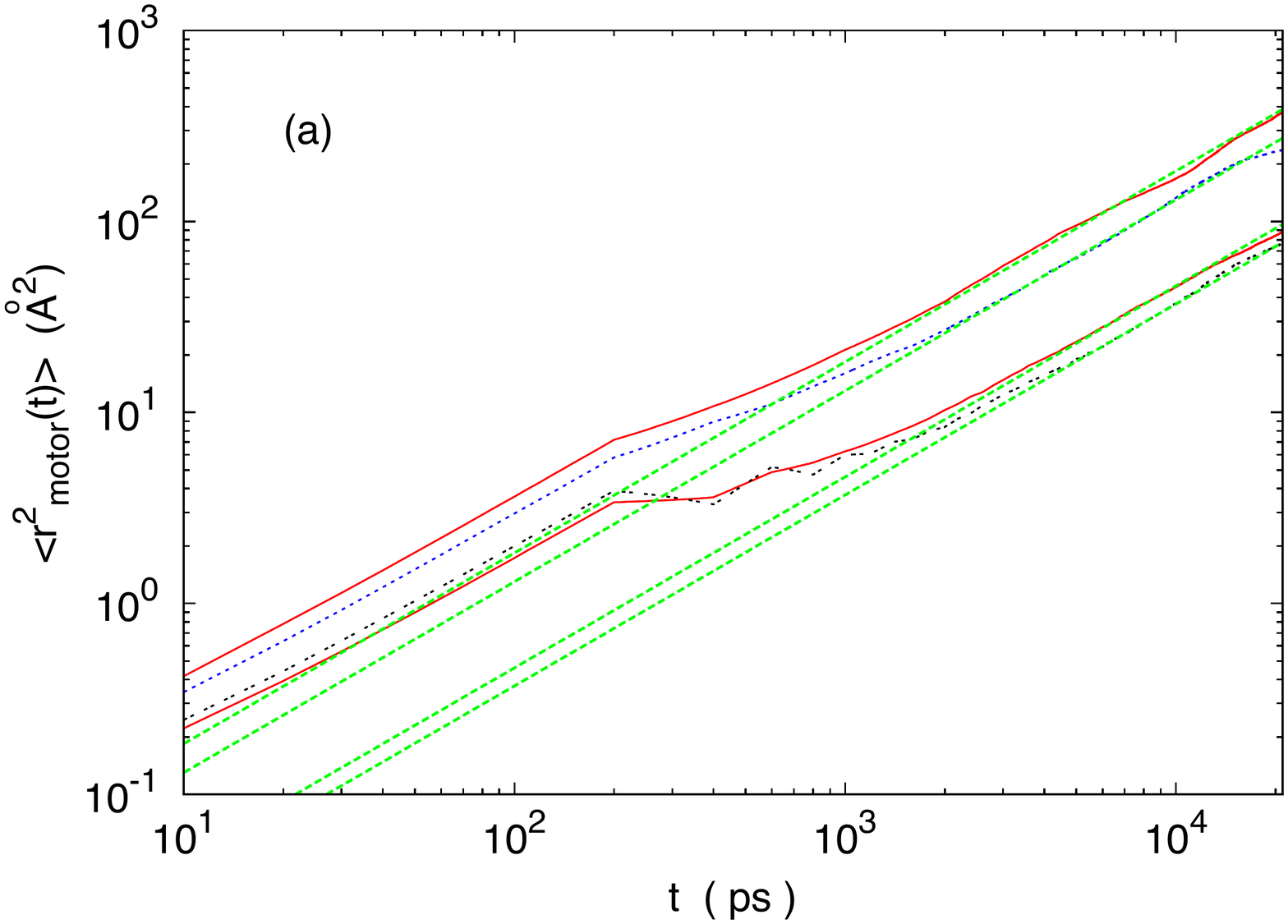}

\includegraphics[height=7cm]{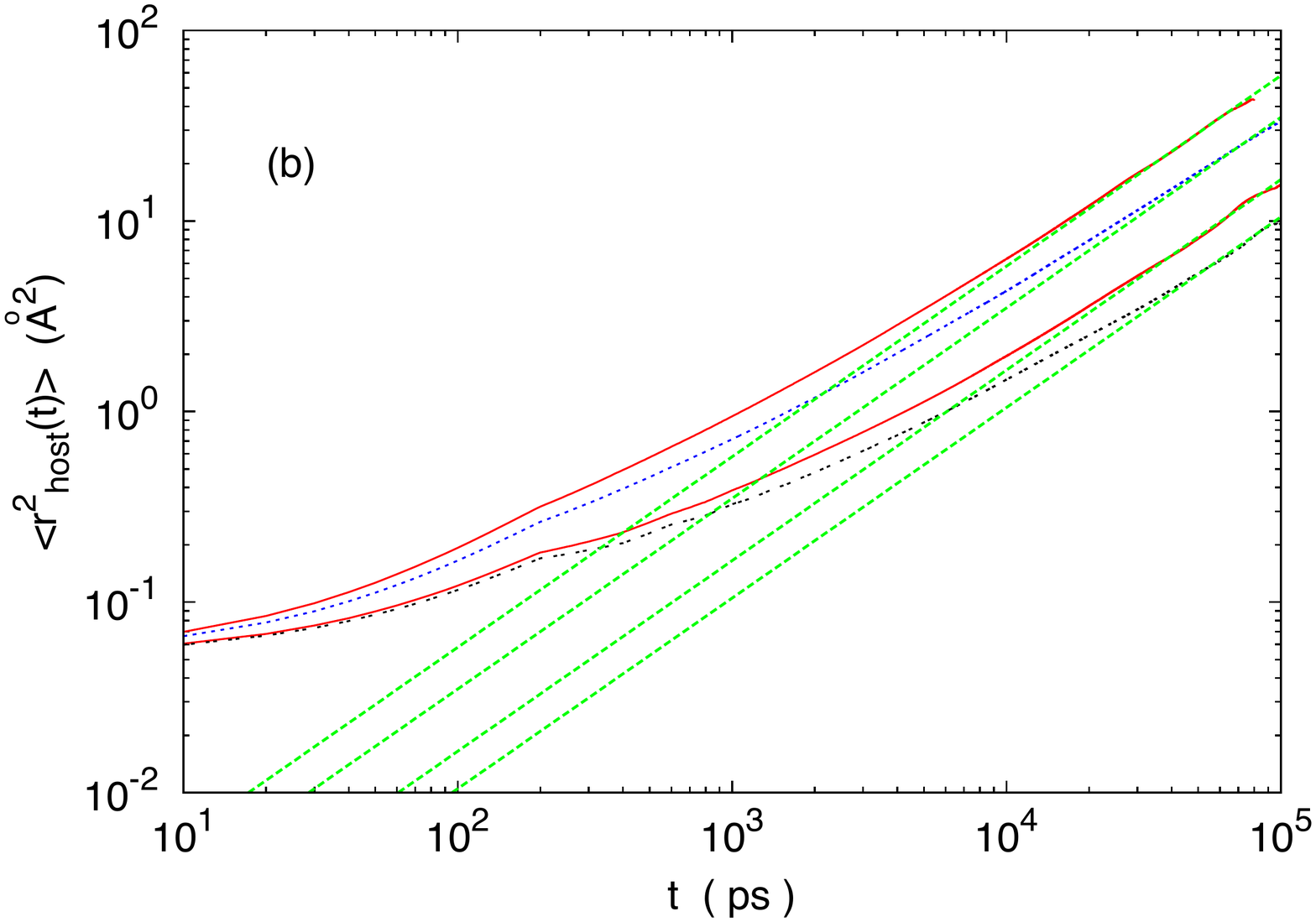}

\caption{(color online) Mean square displacement of the motor (a) and host (b) for various folding times $\tau_{f}$.
From top to bottom: $\tau_{f}=3$ $10^{-2} ps$,  $4$ $10^{-2}ps$,  $2$ $10^{-1}ps$, and $2$ $ps$. The dashed lines are linear fits showing that the mean square displacement is proportional to $t$ at large time scale.
 } 

 \label{f9}
\end{figure}

In our study, for the calculation of he diffusion coefficients, we have supposed that the motor's and host motions were diffusive. 
We will show here that it is actually the case i.e. that  the mean square displacement displays a linear dependence on time at large time scales.
For that purpose we show in Figure \ref{f9} the motor and host's mean square displacements time dependence for various folding times $\tau_{f}$ ranging from $\tau_{f}=3$ $10^{-2} ps$ to $2$ $ps$. 
The fits show that the mean square displacements evolve linearly with time at large time scales.
Consequently the motions are diffusive.
Note that diffusive motions are expected as the motor has no preferential direction of motion in the amorphous environment of the study and the motion is Markovian on times larger than $T_{f}$.

\begin{thebibliography}{99}




\bibitem{motor1} R.D. Astumian, 
 \newblock \emph{Science }  {\bf 276}, 917-922 (1997).

\bibitem{motor2} M.F. Hawthorne, J.I. Zink, J.M. Skelton, M.J. Bayer, C. Liu, E. Livshits, R. Baer, D. Neuhauser, 
 \newblock \emph{Science }  {\bf 303}, 1849-1851 (2004).

\bibitem{motor3} P. Palffy-Muhoray, T. Kosa, E. Weinan, 
 \newblock \emph{Appl. Phys. A} {\bf 75}, 293-300 (2002).

\bibitem{motor4} J. Berna, D.A. Leigh, M. Lubomska, S.M. Mendoza, E.M. Perez, P. Rudolf, G. Teobaldi, F. Zerbetto, 
 \newblock \emph{Nature Mater.}  {\bf 4}, 704-710 (2005).

\bibitem{motor5} T.R. Kline, W.F. Paxton, T.E. Mallouk, S. Ayusman,
 \newblock \emph{Angew. Chem. Int. Ed.}  {\bf 44}, 744-746 (2005).

\bibitem{motor6} W.R. Browne, B.L. Feringa, 
 \newblock \emph{Nature Nanotech.}  {\bf 1}, 25-35 (2006).

\bibitem{motor7} K. Dholakia, P. Reece, 
 \newblock \emph{Nanotoday}  {\bf 1}, 20-27 (2006).


\bibitem{motor8} T. Fehrentz, M. Schonberger, D. Trauner, 
 \newblock \emph{Angew. Chem. Int. Ed.} {\bf 50}, 12156-12182 (2011).


\bibitem{motor10} M.M. Russew, S. Hecht,
 \newblock \emph{Adv. Mater.} {\bf 22}, 3348-3360 (2010).

\bibitem{motor11} N. Katsonis, M. Lubomska, M.M. Pollard, B.L. Fearing, P. Rudolf,
 \newblock \emph{Progress in Surface Science}  {\bf 82}, 407-434 (2007).


\bibitem{motor12} A.P. Davis,
 \newblock \emph{Nature}  {\bf 401}, 120-121 (1999).

\bibitem{motor13} J.P. Sauvage,
 \newblock \emph{Molecular machines and motors} , {Springer, Berlin},{ 2001}.


\bibitem{motor14} E.R. Kay, D.A. Leigh,
 \newblock \emph{Nature}  {\bf 440}, 286-287 (2006).


\bibitem{motor15} V. Balzani, et al.,
 \newblock \emph{Proc. Nat. Acad. Sci.}  {\bf 103}, 1178-1183 (2006).


\bibitem{motor16} T. Muraoka, K. Kinbara, Y. Kobayashi, T. Aida,
 \newblock \emph{JACS}  {\bf 125}, 5612-5613 (2003).


\bibitem{motor17} T.J. Huang, et al.,
 \newblock \emph{Nano Lett.}  {\bf 4}, 2065-2071 (2004).


 \bibitem{pccp}  S. Ciobotarescu, N. Hurduc, V.Teboul
\newblock  {\em Phys. Chem. Chem. Phys.}  {\bf  14}, 5699-5709 (2016).


\bibitem{anderson}  
 P.W. Anderson,
\newblock  {Science}  { \bf  267}, 1610 (1995).


\bibitem{gt1} K. Binder, W.  Kob, 
\newblock  {\em  Glassy Materials and Disordered Solids}, World Scientific, Singapore 2011


\bibitem{gt4} P.G. Debenedetti, 
\newblock  {\em  Metastable Liquids}, Princeton Univ. Press, Princeton 1996

\bibitem{gt2} P.G. Wolynes,  V.  Lubchenko, 
\newblock  {\em  Structural Glasses and Supercooled Liquids}, Wiley, Hoboken 2012

\bibitem{adam} G. Adam, J.H. Gibbs, 
\newblock  {J. Chem. Phys.} {\bf 43}, 139 (1965)


\bibitem{chandler}  J.P. Garrahan, D. Chandler,
\newblock  {Phys.Rev. Lett.}  { \bf 89}, 035704 (2002).



\bibitem{frust}  
 G. Tarjus, S.A. Kivelson, Z. Nussimov, P. Viot, J. Phys.:Condens. Matter { \bf 17}, R1143 (2005) .

\bibitem{frust2}  
 D. Kivelson, S.A. Kivelson, X.L. Zhao, Z. Nussimov, G. Tarjus, Physica A { \bf 219}, 27 (1995).

\bibitem{RFOT}  
T.R. Kirkpatrick, D. Thirumalai, P.G. Wolynes, Phys.Rev. A { \bf40}, 1045 (1989).


\bibitem{RFOT2}  
X.Y. Xia, P.G. Wolynes, PNAS { \bf 97},  2990 (2000).

\bibitem{MCT}  
W. Gotze, Complex Dynamics of Glass-Forming Liquids: A mode-coupling theory (Oxford University Press, Oxford 2008).




\bibitem{defect}  
F. Ritort, P. Sollich, Adv. Phys. {\bf 52},  219 (2003).

\bibitem{facilitation}  
D. Chandler, J.P. Garrahan, Annu. Rev. Phys. Chem. {\bf 61}, 191 (2010).

\bibitem{dh1} L. Berthier, G.  Biroli, J.P. Bouchaud, L. Cipelletti, W. Van Saarlos, 
\newblock  {\em  Dynamical Heterogeneities in Glasses, Colloids and Granular Media}, Oxford Univ. Press, Oxford 2011

\bibitem{dh2}  M. Vogel, S.C. Glotzer, 
\newblock  {Phys. Rev. E}  { \bf 70}, 061504 (2004).

\bibitem{dh3}  M. Vogel,  S.C. Glotzer, 
\newblock  {Phys.Rev. Lett.}  { \bf 92}, 255901 (2004).

\bibitem{dh4}  V. Teboul, A. Monteil, L.C. Fai, A. Kerrache, S. Maabou, 
\newblock  {Eur. Phys. J. B}  { \bf 40}, 49-54 (2004).

\bibitem{dh5}  Z. Zheng, R. Ni, F. Wang, M. Dijkstra, Y. Wang, Y. Han,
\newblock  {Nature Comm.}  { \bf 5}, 3829 (2014).



\bibitem{fragile1}  C.A. Angell, 
\newblock  {Science}  { \bf 267}, 1924-1935 (1995).


\bibitem{fragile2}  N.A. Mauro, M. Blodgett, M.L. Johnson, A.J. Vogt, K.F. Kelton, 
\newblock  {Nature Comm.}  { \bf 5}, 4616 (2014).


\bibitem{prl}  V. Teboul, M. Saiddine, J.M. Nunzi, 
\newblock  {\em Phys. Rev. Lett. } {\bf  103}, 265701 (2009).



\bibitem{us2}  
 V. Teboul, J.B. Accary, M. Chrysos,
\newblock  {Phys. Rev. E}  {\bf 87},  032309 (2013).



\bibitem{c1}  
 R. Tavarone, P. Charbonneau, H. Stark,
\newblock  {J. Chem. Phys.}  {\bf 144}   104703 (2016).


\bibitem{c2}  
 D. Orsi, L. Cristofolini, M.P. Fontana, E. Pontecorvo, C. Caronna, A. Fluerasu, F. Zontone, A. Madsen, 
\newblock  {Phys. Rev. E}  { \bf 82}   031804 (2010).



\bibitem{c3}  
 T. Mori, S. Saito,
\newblock  {J. Chem. Phys.}  {\bf 142}   135101 (2015).





\bibitem{dif1} P. Karageorgiev, D. Neher, B. Schulz, B. Stiller, U. Pietsch, M. Giersig, L. Brehmer, 
\emph{Nature Mater.}  {\bf4}, 699-703 (2005).

\bibitem{dif2} G.J. Fang, J.E. Maclennan, Y. Yi, M.A. Glaser, M. Farrow, E. Korblova, D.M. Walba, T.E. Furtak, N.A. Clark, 
\emph{Nature Comm.}  {\bf4}, 1521 (2013).

\bibitem{dif3} N. Hurduc, B.C. Donose, A. Macovei, C. Paius, C. Ibanescu, D. Scutaru, M. Hamel, N. Branza-Nichita, L. Rocha,
\emph{Soft Mat.}  {\bf 10}, 4640-4647 (2014).

\bibitem{dif4}  J. Vapaavuori, A. Laventure, C.G. Bazuin, O. Lebel, C. Pellerin, 
\emph{J. Amer. Chem. Soc.}  {\bf 137}, 13510 (2015).



\bibitem{cage}  V. Teboul, M. Saiddine, J.M. Nunzi, J.B. Accary, 
 \newblock \emph{J. Chem. Phys. }  {\bf134}, 114517 (2011).
 
\bibitem{c5}  
D. Bedrov, J.B. Hooper, M.A. Glaser, N.A. Clark,
\newblock  {Langmuir}  { \bf   32}, 4004 (2016).





\bibitem{review} A. Natansohn,  P. Rochon, 
 \newblock \emph{Chem. Rev. } {\bf102}, 4139-4175 (2002).
 

\bibitem{review2} J.A. Delaire, K. Nakatani, 
 \newblock \emph{ Chem. Rev.} {\bf 100}, 1817 (2000).
 

\bibitem{review4} K.G. Yager, C.J. Barrett,  
\newblock \emph{Curr. Opin. Solid State Mater. Sci.} {\bf 5}, 487 (2001).



\bibitem{a17}  T.G. Pedersen, P.M.  Johansen, 
\newblock  {\em Phys. Rev. Lett.}  {\bf  79}, 2470-2473 (1997).

\bibitem{a18}  T.G. Pedersen, P.M.  Johansen, N.C.R.  Holme, P.S.  Ramanujam, S. Hvilsted,
\newblock  {\em Phys. Rev. Lett.}  {\bf  80}, 89-92 (1998).

\bibitem{a21}  J. Kumar, L.  Li, X.L.  Jiang, D.Y.  Kim, T.S.  Lee, S. Tripathy, 
\newblock  {\em Appl. Phys. Lett.}  {\bf  72}, 2096-2098 (1998).



\bibitem{a22}  C.J. Barrett, P.L.  Rochon, A.L.  Natansohn, 
\newblock  {\em J. Chem. Phys.}  {\bf  109}, 1505-1516 (1998).

\bibitem{a23}   C.J. Barrett, A.L. Natansohn, P.L. Rochon, 
\newblock  {\em J. Phys. Chem.} {\bf  100}, 8836-8842 (1996).

\bibitem{a24}  P. Lefin, C.  Fiorini,  J.M. Nunzi, 
\newblock  {\em Pure Appl. Opt.}  {\bf  7}, 71-82 (1998).








 \bibitem{rate}   J.B. Accary, V. Teboul, 
\emph{J. Chem. Phys.}  {\bf139}, 034501 (2013).


\bibitem{scallop1} E. M. Purcell,
 \newblock \emph{Am. J. Phys.} {\bf 45}, 3 (1977)

\bibitem{scallop2}  T. Qiu, et al.,
 \newblock \emph{Nature Comm.} {\bf 5}, 5119 (2014)

\bibitem{scallop3} E. Lauga,
 \newblock \emph{Phys.Rev.Lett.} {\bf 106}, 178101 (2011)



\bibitem{Crooks} E.H. Feng, G.E. Crooks, 
 \newblock \emph{Phys.Rev.Lett.} {\bf 101}, 090602 (2008)

\bibitem{Crooks1}  G.E. Crooks, 
 \newblock \emph{J. Stat. Phys.} {\bf 90}, 1481 (1997)

\bibitem{Crooks2} G. M. Wang, E. M. Sevick, E. Mittag, D. J. Searles, D. J. Evans,
 \newblock \emph{Phys.Rev.Lett.} {\bf 89}, 050601 (2002)

\bibitem{Crooks3} D. Collin, F. Ritort, C. Jarzynski, S. B. Smith, I. Tinoco Jr, C. Bustamante,
 \newblock \emph{Nature} {\bf 437}, 231 (2005)


\bibitem{mdcm-1}  
E. Flenner, G. Szamel,
\newblock  {J. Phys. Chem. B}  { \bf  119}, 9188 (2015).

\bibitem{mdcm-2}  
E. Flenner, G. Szamel,
\newblock  {Nature Comm.}  { \bf  6}, 7392 (2015).

\bibitem{mdcm-3}  
Y.S. Elmatad, A.S. Keys,
\newblock  {Phys. Rev. E}  { \bf  85}, 061502 (2012).

\bibitem{md7}  V. Teboul, Y. Le Duff,
 \newblock  {\em  J. Chem. Phys.}  {\bf 107}, 10415-10419 (1997).

\bibitem{mdcm-4}  
B. Uralcan, I.A. Aksay, P.G. Debenedetti, D. Limmer,
\newblock  {J. Phys. Chem. Lett.}  { \bf  7}, 2733 (2016).

\bibitem{mdcm-5}  
J. Zhang, F. Muller-Plathe, F. Leroy,
\newblock  {Langmuir}  { \bf  31}, 7544 (2015).





 \bibitem{angle}  V. Teboul,
\newblock  {\em J. Phys. Chem. B}  {\bf  119}, 3854 (2015).



\bibitem{md4}  
J.B. Accary, V. Teboul, 
\newblock  {\em J. Chem. Phys.} {\bf  136}, 0194502 (2012).



\bibitem{allen} M.P. Allen,   D.J. Tildesley, 
\newblock  {\em  Computer Simulation of Liquids}, Oxford University Press, New York 1990

\bibitem{md2} M. Griebel, S. Knapek, G. Zumbusch, 
\newblock  {\em  Numerical Simulation in Molecular Dynamics}, Springer-Verlag, Berlin 2007

\bibitem{md2b} D. Frenkel, B. Smit, 
\newblock  {\em  Understanding Molecular Simulation}, Academic Press, San Diego 1996


\bibitem{berendsen}  H.J.C. Berendsen,  J.P.M. Postma, W. Van Gunsteren,  A. DiNola, J.R.  Haak, 
 \newblock  {\em  J. Chem. Phys.}  {\bf 81}, 3684-3690 (1984).






\bibitem{David} D. Chandler, 
\newblock  {\em  Introduction to Modern Statistical Mechanics}, Oxford University Press, Oxford 1987





\end{thebibliography}
\end{document}